# Legal Matters in Research Software: A Few Things Worth Discussing

## Giuditta Parolini *

giuditta.parolini@mfn.berlin

Museum für Naturkunde - Leibniz Institute for Evolution and Biodiversity Science
Berlin, Germany

**Abstract:** The paper discusses legal aspects relevant to the development of research software and practical approaches taken by research software engineers to deal with them. Intellectual Property Rights on software are considered alongside licensing choices made by the research community. The discussion addresses the ambiguities in the identification of the copyright holder of research software, the uncertainty surrounding liability, and remarks the varying level of support on legal matters provided by research organisations. The paper also reflects on the widespread use of AI coding assistants in the absence of institutional policies, and on the new AI regulations passed by the European Union. The aim of the contribution is to point out that a better understanding of legal matters concerning software development is an asset in giving research software the right value it deserves as a driver of scientific progress.

**Keywords:** Copyright, Research software, Software licences, AI coding assistants, AI Act

## 1 Introduction

Developing research software requires much more than writing code. Quality assurance, creation of appropriate documentation for both developers and end users, licensing, publishing are equally important aspects in developing software that can be trusted by the research community, shared for reuse, and integrated in other software projects. However, licensing—and more generally all the legal aspects associated with the making of research software—are often an afterthought in the process. This is not surprising given that curricula teaching software development hardly touch upon the subject and the average software developer struggles with licence compatibility issues involving even popular open-source licences [AMWH19]. Even if the subject had a greater relevance in the computer science curriculum, the problem would not be solved because the majority of research software engineers are not formally trained in computer science, according to a recent survey of the field [CMK22].

This situation is rather unfortunate, because legal aspects matter at all stages of research software development and have a direct impact on licensing choices. Who is the copyright holder of research software? Who is liable when there are issues with a piece of research software? Have all third-party Intellectual Property Rights (IPR) associated with the use of external software libraries being properly considered? Where does the research software engineering community

* The author is sponsored by the German Research Foundation (DFG) under Grant Number 528674292.

position itself in relation to FLOSS (Free/Libre and Open Source Software) and proprietary software? The recent popularity acquired by AI coding assistants is further increasing the legal complexity associated with developing research software, as it has an impact on IPR and poses a potential risk of infringing third-party software licences. For organisations based in the European Union, newly passed legislations, such as the AI Act, also require to assess practices in research software and data management to ensure that they fully comply with the current legislative framework.

The Conference of the German Research Software Engineering Association (deRSE) held in Karlsruhe in February 2025 provided an opportunity to discuss these legal aspects and gather an overview of how research software engineers deal with them in practice. The present paper is inspired by the discussion with the workshop participants. It touches upon the ambiguities in the identification of the copyright holders of research software and the uncertainty surrounding liability. It remarks the varying level of legal support in software licensing provided by research organisations, but also reflects on the widespread use of AI coding assistants and its meaning for research software in the absence of institutional policies and regulations. The paper also considers the new AI legislation passed by the European Union and its consequences for research organisations. Without making any claim to statistical representativeness—the workshop participants were a self-selected group of conference attendees—the responses received and the discussion on the day offer some points worth further consideration.[1]

Every contribution about legal matters always has a disclaimer that information provided should not be interpreted as legal advice. Also this paper does not wish to provide any legal advice. The aim of the paper is rather to point out that a better understanding of legal matters is an asset in giving research software the right value it deserves as a driver of scientific progress.

## 2   The Legal Status of Research Software

### 2.1   Software Intellectual Property Rights

IPR tools relevant to software are copyright, patents, trade secrets, and trademarks. Copyright protects original work created by one or more individuals and fixed in a tangible medium of expression, such as a book, a musical score, or a computer script. The patenting of software (see Section 2.5), instead, has been "a debatable subject matter" [Abr09] with significant differences among national legislations [Lau06] and openly opposed by the FLOSS community [Sta02]. For a specified period of time, patents prevent others from making (even via reverse engineering), using, or selling the idea behind the protected invention without explicit permission—usually given in exchange for payment—of the patent holder. Trade secrets are the IPR that protect confidential information. They do not need to be registered and IPR protection continues until the confidential information becomes public. In the case of software, an algorithm or the entire source code can be considered a trade secret and protected as a trade secret. Trademarks are the IPR that covers commercial names of products and services, their logos, design, etc. In the case of software, trademarks can be the software name and its logo, for instance. Trademarks are often officially registered. Copyright and patents are the IPR tools generally used by research organisations and they are discussed in more detail below.

---

[1] A list of the workshop questions and answers is available in Appendix. Please note that the participants' answers were always voluntary, therefore there may be a different total number of responses to each question.

## 2.2 Software Copyright

Under copyright laws, software is protected as a "literary work" (for the European Union, see [euc91] and [eus09]). The copyright holder has both economic and moral rights, according to the Berne Convention for the Protection of Literary and Artistic Works [ber79]. That is to say, the copyright holder is entitled to receive the financial benefits that may be associated to the creative work protected by copyright (economic rights), and the copyright holder has also the right to be mentioned as the author of the work protected by copyright (moral rights). Copyright protection begins with the creation of the work without need of any registration/application. In the European Union, copyright protection continues until seventy years after the death of the copyright holder or the last surviving copyright holder for works created by multiple authors. The long-term span of copyright protection explains why almost all the software so far produced, unless it was originally released in the public domain, is still under copyright.[2]

In the European Union, originality requirements for software are minimal. According to the EU Directive on the Legal Protection of Computer Programs, the only criterion for software copyright protection is originality "in the sense that it is the author's own intellectual creation" [eus09]. Member states cannot impose other criteria, as this could create paradoxical situations in which a piece of software receives copyright protection in a country of the European Union and not in another. As a result of this legislation, any piece of software—commercial, non-commercial, written by amateurs or professionals, in any stage of the software lifecycle—enjoys copyright protection as long as it is produced by an author. The author can only be a natural person, according to current copyright law. If tools, such as robots, AI models, algorithms, are involved in code development, there must be a demonstrated human contribution to make the work copyrightable. The copyright holder has a right of choice, protected by law, about how the software developed should be distributed. Here the software licences discussed in Section 3 come into play.

## 2.3 The Copyright Holder(s) of Research Software

Software copyright can be transferred from its original author to a third-party, such as a company, an institution, another natural person, using a contractual agreement. Contractual agreements typically establish that the copyright on the code developed by staff or external contractors hired by a business company is owned by the company and not by the developers writing the code. When it comes to research software the situation is often complex and the answer changes case-by-case, as the majority of the workshop participants in Karlsruhe confirmed.

One of the issues is that research software is developed by people who work in different positions within research organisations. Some may have a specific research software engineering role and a contract that makes provisions about code copyright ownership from the start, but in many cases research software is also developed by people working in research roles, both postdocs on short-term contracts and staff on permanent research and teaching positions. IPR on research output is regulated differently in different countries for researchers. For instance, work done by employees of U.S. federal research organisations as part of their employment duties is typically released in the public domain. This does not apply to staff working in non-federal U.S. research organisations. Indeed, in many countries researchers are the copyright owners of their intellectual work. As a matter of fact, it is not the scientific institution, but the researcher who signs a copyright transfer agreement with publishers when a journal accepts their scientific articles. In these cases, if there is code, for instance a data analysis script, associated to the scientific publication, how should this be considered? Can a researcher own the copyright on the paper, but not on the software that has been instrumental in conceiving the paper results? Who is in charge, then, of licensing and publishing the code to ensure reproducibility of the scientific results? Different considerations may apply to large research software projects, where a software package or a web application is the only expected output. Some research institutions have provisions for these cases, especially if a commercial interest is envisioned. However, large part of the research software projects are very valuable but not in the commercial sense, and IPR in these cases often remains unclear or it is decided on a case-by-case basis.

---

[2] The word *software* was first used in print only in 1958 [Tuk58].

A further complicating factor is posed by student contributions to research software. Students may contribute to write code for a research project as part of their studies and without being in any contractual relationship with the research organisation. Unless some agreement is formalised and signed, they maintain the copyright on the code they wrote and managing the software IPR associated to their contributions becomes problematic.

## 2.4 IPR in Software Developed by Community Projects

Unclear copyright ownership becomes a very important issue when the software project is ambitious, involves multiple contributors from different institutions, and the resulting output is expected to be beneficial for a large scientific community. The way open-source communities manage copyright issues in the development and maintenance of large software projects may offer some food for thoughts for research software. Several well-established open-source projects in the data science ecosystem, such as NumPy, Pandas, and Scikit-Learn, use a U.S. public charity, NumFOCUS, as a fiscal sponsor. Having a third-party as a legal entity representing the software project simplifies several administrative and legal issues, including IPR, as this arrangement guarantees that "project assets are owned by the project and not individual contributors".[3] Interestingly, one of the talks presented at the conference in Karlsruhe, discussed the transition of a research software project under the NumFOCUS umbrella [KCH+25]. The transition required time and multiple copyright transfers. The software copyright rested initially with the research institutions where the project had started sponsored by third-party grants. Therefore, the academic institutions had to agree to transfer the copyright back to the developers who, in turn, could then transfer the software copyright of the project to NumFOCUS. Passing on administrative and legal issues—not only IPR, but also the liability that is always associated to distributing software—to a non-profit organisation is a strategy often employed by successful research software. Another example that can be cited and that was discussed with the workshop participants is Zotero, an open-source citation manager quite popular in academia, especially in the humanities. Zotero was born as a research software project at George Mason University in Virginia in the early 2000s, but has since transitioned as a project under the umbrella of the Corporation for Digital Scholarship, a U.S. non-profit organisation that manages several large digital humanities software projects.

---

[3] https://numfocus.org/projects-overview

In its guidelines related to the handling of research software [Deu24], the German Research Foundation explicitly addresses rights and authorship as something that "[b]efore starting a research project, [...] must be clarified and, if necessary, contractually regulated with employees, students or external service providers" [p. 11]. The examples discussed above suggest that having such agreements in place is essential to continue software projects beyond the limited time-frame of the initial grant assigned to a research group or an institution. It is also beneficial for the contributors to have an agreement that gives them a clear indication of how their intellectual property will be handled by the project and makes provisions for the liability associated to it. Perhaps, it would be helpful if research funders were to take a more active role in advising and supporting researchers and institutions in addressing these legal matters related to research software on a practical level. The initial effort would pay back in terms of the overall life perspective of the software project. Clarity about IPR is not the only, but certainly a main requirement, to help research software grow into a long-term project with an active group of developers, maintainers, and users.

## 2.5 Software Patents

Patents are the other IPR protection on research software worth mentioning. The European Patent Convention states that patents "shall be granted for any inventions, in all fields of technology, provided that they are new, involve an inventive step and are susceptible of industrial application" (Article 52(1)) [eup20]. The Convention, however, explicitly excludes "discoveries, scientific theories and mathematical methods", but also "programs for computers" and a handful of other items from the list of patentable inventions (Article 52(2)). This exclusion is not absolute, but applies only if the invention is claimed "as such" (Article 52(3)). Therefore, patents involving software can be granted by the European Patent Office and by the patent offices of the member states of the European Union. The European Patent Office, for instance, can grant a patent for a *computer-implemented invention* as long as the invention has a technical character. Computer-implemented invention is an "expression intended to cover claims which involve computers, computer networks or other programmable apparatus whereby prima facie one or more of the features of the claimed invention are realised by means of a [computer] program or programs" [EK21][p.8]. Patentable computer-implemented inventions range from a technical process for improving digital image processing (e.g., the VICOM case in the 1980s) to computer-implemented methods of determining oxygen saturation in blood [Eur25][Chapter IV], just to mention a few examples. Unlike copyright, an application need to be filed for receiving a patent. Patents actively prohibit reverse engineering while software only protected by copyright can be reverse engineered without any legal infringement of copyright. Technology transfer offices at research institutions might evaluate whether a piece of software is worth patenting, i.e., can have a commercial value, and take care of the process. Patent protection, once granted, is much more limited in time (twenty years) compared to copyright protection.

There are cases in which research organisations have resorted to patents for protecting their software IPR and have exploited it commercially. The patents on mp3 software held by the Fraunhofer Institute for Integrated Circuits are a case likely well known to many German readers.[4] Aside from very specific cases in which research software has business relevance, however, patents are not a standard tool to manage research software IPR. Patenting is a long and costly business that research organisations pursue only in selected cases.

As mentioned in Section 2.1, the FLOSS community takes a critical stance towards software patents and has openly expressed concerns about the risk that patents may pose for developers and users of FLOSS software [MCB14]. The Open Source Initiative, a non-profit organisation that advocates for open source and maintains a catalogue of software licences compliant with the Open Source Definition, rejects all licences that explicitly refuse to grant a patent licence alongside the copyright licence [Ope23]. The rationale behind the choice is that such licences fail to meet key requirements of the Open Source Definition, such as redistributing the software without need of additional licences (Art. 7). The Open Source Initiative, however, accepts licences that do not have a specific patent clause, if such a clause is implied by the formulation of the licence. Therefore, currently, FLOSS software licences may or may not include an explicit patent clause (see Section 3). The Free/Libre Software community has explicitly included a patent retaliation clause in version 3 of the GPL licence (GPL v.3). Due to this clause, software users and developers are prevented from filing patent suits against other software licencees and would have their own licence revoked if they attempted to do so.

## 3  Licensing Research Software

### 3.1  Reasons for Licensing Research Software

As discussed above, software is protected by copyright, and, in some cases, it is also patented. Therefore, adding a licence to software is the only way to legally enable other people to use, modify, and redistribute software without infringing the rights of the IPR holder. Without a licence, even code that is publicly available on the internet cannot be used by others. The popular developer platform GitHub, which hosts many software development projects including those of research institutions, clearly reminds this to its users inviting them to make an "informed decision" about how to license their repository and emphasising that "showing" your code to others is very different from legally enabling them to reuse it.[5] Ten years ago (2015) only 20% of GitHub repositories had a licence (ca. 30% if forked repositories are included in the count). Since then, the number has grown to some extent, but according to a 2020 analysis that only focused on the most popular GitHub repositories, almost half of these repositories still does not have a licence.[6] The absence of a licence might not be a crucial issue for personal projects and portfolios hosted on GitHub—even though a licence would certainly signal that the project author has a clear understanding of the overall framework surrounding software development and not just of its technical aspects— but it becomes critical when code is not just meant to "showcase" technical competence. Anyone reusing code published on the Internet without a proper licence is committing copyright infringement and can potentially be sued for this infringement. As a result, unlicensed research software does not satisfy in any way the key requirement of re-usability stated in the FAIR principles for research software [CKB+22], and conflicts with any *open science* [BEG+03] and *open knowledge* [Ope17] aspiration that research institutions may claim to have. Therefore, all research organisations should offer training and advice on licences to staff developing software. The responses gathered during the workshop suggest that this is not yet the case. Almost 70% of the participants answering during the workshop stated that their institutions do not provide any training or advice regarding software licences.

---

[4] The Fraunhofer Institute for Integrated Circuits held several software patents on the mp3 audio format and benefited from the related royalties until 2017, when the last mp3 patent expired [Ste14].

[5] See GitHub user documentation and the GitHub blog post "Open source license usage on GitHub.com" (2015, updated 2021).

[6] See this blog post. Given that the 2015 analysis was done by GitHub and considered all GitHub repositories, while the most recent was carried out by a SaaS company and only took into account popular repositories, the percentage of unlicensed code on GitHub is likely to be much higher than 50% even today.

## 3.2 Licences for Research Software

A software licence makes provisions for the acceptable use of the software and the restrictions to it, sets the conditions for sub-licensing and redistributing, and includes clauses stating the limitation of liability and the extent to which a warranty is accepted or disclaimed. Licences may also have provisions for software patents, as it is the case with the Apache 2.0 licence, which is a popular open-source software licence. Proprietary software with a commercial value is often distributed using custom licences that describe in great detail the customer limitations in terms of software use, such as the length of time for which the licence is granted and the maximum number of devices on which the software can be installed. Typically, proprietary software cannot be redistributed but, if this is allowed in any form, the licence describes all the conditions that need to be met to redistribute the software.

Research software is usually distributed under a FLOSS licence, as confirmed by the answers gathered from the workshop participants in Karlsruhe. Both permissive open-source licences, such as the MIT licence, and copyleft licences, like the GPL licence, are used (See Section 3.3). In the experience of the workshop participants the research institution usually does not influence the choice of the software licence. On the other hand, research funders in many cases do. For instance, in the already mentioned Guidelines on Research Software the German Research Foundation recommends "the greatest possible [...] openness of the licence for scientific use" [Deu24][p. 8]. Similarly, recommendations within research projects sponsored by the European Union invite to always make research software available under an open-source licence [Eur20].

Both proprietary and open-source software licences are enforceable, that is to say it is possible to legally act in court when the terms of the licence are not respected. Germany, for instance, is a country with numerous legal cases related to the enforcement of open-source software licences due to the engagement of an activist, Harald Welte, who has legally pursued violations of the GPL licences in court [Bro22].

An important, but often overlooked point, is that adding a licence to research software does not only allow legal reuse, but also protects the author/copyright holder from legal consequences. Whenever software is created and distributed there is an associated liability, i.e., potential legal consequences, for instance, if the software malfunctions and causes damages to the user, or if the software does not fulfil security standards. Proprietary software licences deal at length with liability and user warranty to limit as much as possible the responsibility of the software manufacturer. All FLOSS licences have a disclaimer of liability as well, because even software that is distributed free of charge is equally subject to liability. For instance, in the MIT licence, which is the most popular open-source licence [7], the liability disclaimer represents almost half of the entire text and begins with "THE SOFTWARE IS PROVIDED "AS IS", WITHOUT WARRANTY OF ANY KIND ..." (see Figure 1). Liability clauses are often limited, in practice, by national legislations that may not deem them enforceable in case of gross negligence or wilful misconduct. However, they remain a key element in managing risk while developing software because software is never completely bug-free and always requires updates and improvements [Goe16].

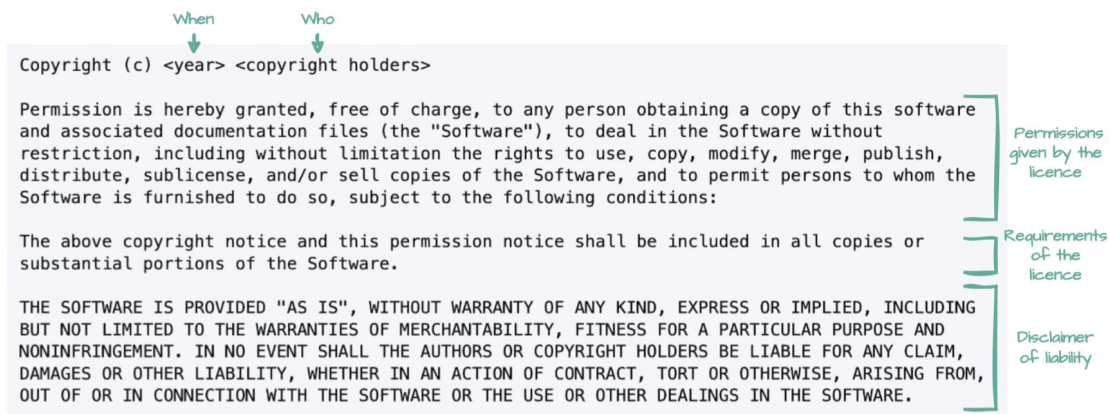

**Figure 1: The MIT licence**
The disclaimer of liability is written all in capital. This is common in open-source software licences, as the disclaimer must be conspicuous.

The directive on product liability recently approved by the European Parliament [eup24] concerns also software and increases the liability of software developers. Developers who market their software products in the European Union can now be held responsible for insufficient cybersecurity and explicitly called to answer for data loss or non-pecuniary damages. FLOSS software that is not commercially developed does not fall within the scope of the directive unless there is some form of payment involved or personal data are collected in exchange for software use [eup24] [Art. 14]. Research software engineers should be aware of the additional liability they may face in the European Union when the directive becomes part of national legislations and should carry out risk assessments for their software with even greater diligence.

A last point worth mentioning is the existence of the dual licensing option that is sometimes used by research institutions to comply with open science principles and to exploit commercially their research software at the same time. Dual licensing is only possible when the institution or its research software engineer(s) are the only IPR owners and means redistributing the software under two separate licences, typically, a strong copyleft licence, such as the GPL licence, and a commercial licence that allows the licensee to use the software within a proprietary code base without need to redistribute the source code. One example of research software that uses the dual licensing model is the C-library FFTW (Fastest Fourier Transform in the West) for computing the discrete Fourier transform. The library is available both under an open-source licence (GPL v.2) and under a commercial licence with pricing options currently ranging from $2500 to $12500, depending on the library version.

---

[7] For the popularity of FLOSS software licences see GitHub License Rankings for 2024 and the analysis, related to 2023 data, done by the Open Source Initiative considering the popularity of FLOSS licences for different programming languages [Ope].

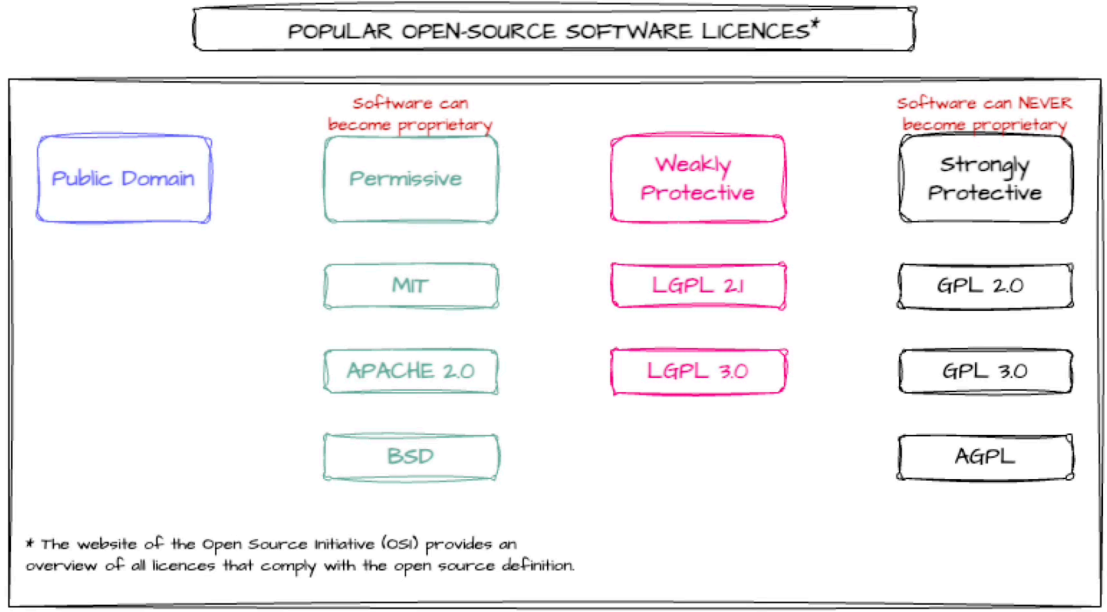

Figure 2: Open-source licences commonly used to distribute software. For a list of OSI-compliant licences see here.

## 3.3 Open-Source Software Licences

Figure 2 lists popular open-source software licences often used to distribute research software. From left to right, Public domain, Permissive, Weakly Protective, and Strongly Protective licences are displayed. Public Domain is in itself not a licence, but it is typically listed in all overviews of software licences because it has been used to freely distribute source code. A popular example is the serverless database SQLite, whose code and documentation have both been dedicated to the public domain by its developers. Public Domain dedication is possible only if the code authors are the only IPR owners. As the Public Domain is not a licence, it does not have a limitation of liability. Permissive software licences, like the already mentioned MIT licence and the Apache 2.0 licence, are so named because they allow to use, modify, and redistribute the source code, but make no requirement about the final licence the modified source code should have. Code distributed under an open-source permissive licence can be reused even within proprietary software and does not require to open the code base to the public. For this reason, open-source permissive licences, in particular the MIT licence, are very popular when source code is distributed. Popular, but for completely different reasons, are also the strongly protective licences, like the GPL licences. They allow to use, modify, and redistribute the source code, but impose to redistribute all code where a GPL library is used under the same original GPL licence to avoid open-source code becoming embedded in proprietary code.[8] For this strong requirement, they are also named *copyleft* licences. The weakly protective licences fit in between the permissive and the strongly protective licences. They still actively prevent open-source software from becoming proprietary, but they include cases where a LGPL library can be used without necessity to licence the entire code base under the same licence.

Understanding software IPR and licensing is important because developers rarely write everything from scratch when they build software. Whenever possible, they rely on pre-existing software libraries. This, however, requires to pay attention to the licences of the individual software pieces and confirm that they are compatible and can be combined before redistributing a software project. For instance, the web-framework Django is distributed under the BSD licence but includes code from the Python standard library, which is licensed under its own special license. This is possible because both Python and Django are distributed under permissive open-source licences. If Python were distributed under a strong copyleft licence, instead, all the Django code base would then need to be released under the same copyleft licence to use code from the Python standard library. The Free Software Foundation has an extensive collection of FAQs on GNU licences that is worth reading for any research software engineer who develops GPL and LGPL software or uses copyleft software. Licence compatibility is a key element when it comes to software redistribution, especially if the software redistributed includes third-party software with different open-source licences. IPR in software redistribution is an intricate legal area that requires careful consideration. Complex cases are probably better discussed with a legal consultant before redistributing source code to avoid infringing third-party IPR.

The responses gathered from the workshop participants in Karlsruhe suggest mixed feelings about the infringement of third-party IPR in the development of research software. The answers that gathered the largest response indicate on the one hand high-confidence in managing and understanding software licences with constant due diligence when third-party software is integrated in the code base, and, on the other hand, uncertainty and lack of support to clarify complex aspects in licence compatibility and software redistribution. This is not surprising given that licence compatibility is a complex topic for all developers [AMWH19] and that many institutions do not provide any support on legal matters related to research software. The situation, however, calls for a more active intervention of research institutions in providing adequate training and, where necessary, grant access to legal advice to avoid possible pitfalls in software licensing and redistribution. In addition, at the beginning of any research software project, it should always be clarified who is liable for infringing third-party IPR, whether the research software engineer or the research institution, and what measures can be taken to minimise the risk of IPR infractions, for instance setting up a clear workflow for tracking the licences of the third-party software libraries used, as described below.

---

[8] A point worth noting, though, is that while GPL v.2 and GPL v.3 licences are not compatible and code released under GPL v.2 cannot be modified and redistributed under GPL v.3, if the code is licensed as *GPL v.2 or later* then derivative work can also be redistributed under the GPL v.3 licence. See this clarification by the Free Software Foundation for more details.

### 3.4 Managing Software Licences

Selecting and managing licences within a software project, especially when the complexity of the project increases, is a non-trivial task but there are a few tools that can help. First of all, *licence selectors*, i.e., tools that allow an overview of the licences available, at least for specific licences. This is the case, for instance, with GitHub Choosealicence for open-source licences. Some licence selectors explicitly guide in the choice of an open-source licence for software and may be helpful for very simple cases if the developer has no previous experience. However, one should always consider that these tools are not meant to provide legal advice and that their suggestions are only as good as the decision tree implemented in the tool. Licence selectors work well for straightforward cases but may not be sufficient when there are multiple factors to account for in the final decision due to the redistribution of third-party libraries. Addressing licence selection early in a research software project can avoid many headaches in the end. As described above, there are clear rules for combining software licences and, once it is clear the outbound licence for redistributing the finished software, these rules help to make informed choices about reusing third-party libraries.

A second category of tools that are helpful to manage licences in software projects are toolkits for licence management. These tools help to streamline and automate the process of managing the licences of the software components used in a project and check that these licences are compatible. Some of these tools are open source, such as Reuse Software, a project of the Free Software Foundation Europe, and Fossology, which is a project of the Linux Foundation. These toolkits are helpful when software projects build on many third-party components and their application in developing research software was explicitly described in one of the talks delivered during the deRSE conference in Karlsruhe.[9] Both toolkits rely on the ISO Standard SPDX (Software Package Data Exchange, ISO/IEC 5962:2021) that defines a "data format for communicating the component and metadata information associated with software packages" [spd21][p. 1]. The standard is driven by the community behind the homonymous open-source project, SPDX, hosted by the Linux Foundation. The SPDX project maintains a list of licences commonly used in open-source software and software documentation, and provides them with unique identifiers. In each file of a software project, the tools made available by the SPDX project allow to cite the applicable licence using a single-line statement that is both human and machine readable. In this way licence compatibility checks can be automated. The SPDX standard and tools have been integrated also in the Licensing Assistant made available by the European Union as part of its open licensing strategy.

## 4 AI and Research Software

### 4.1 AI Coding Assistants

In recent years, Large Language Models (LLMs), like ChatGPT, have known increasing popularity. These AI conversational models interact with the user via a prompt in a natural language. Their replies to the user's queries are a probabilistic determination based on the extensive datasets these models have been trained on. The answers generated by LLMs range from correct to completely mistaken, so called hallucinations. AI coding assistants are LLMs used to generate code or perform other tasks relevant to code development (e.g., documentation, debugging) in response to a natural language prompt. Similarly to LLMs, answers produced by AI coding assistants can be correct, correct but inefficient, partially correct or completely wrong. Some AI coding assistants are explicitly trained on coding repositories, like GitHub Copilot, but also generic LLMs, like ChatGPT, can be queried to assist in code generation.

---

[9] Drees T., Feuchter D., Stary T., Winandi A. (2025) "Three Lessons Learned: How RSEs Succeed in License Management", deRSE2025, Karlsruhe. Presentation available here.

AI coding assistants are increasingly used for writing code and documentation in academia. A recent survey of academic postdoc attitudes towards AI tools suggests that over half of the researchers taking up these tools use them for generating, editing, and troubleshooting code [Nor23]. AI coding assistants are also popular among research software engineers. More than half of the respondents in Karlsruhe confirmed to use them regularly or occasionally. ChatGPT and GitHub Copilot proved to be the most used by the respondents. Some workshop attendees also stated that their institutions have paid subscriptions for AI coding assistants. The discussion in Karlsruhe touched upon AI coding assistants because there are legal issues associated with their use. The main legal problems are the copyright status of AI-generated code and the risk to infringe third-party IPR when using AI coding assistants.

The copyright status of AI-generated code is often overlooked, yet it is a crucial aspect when it comes to software licensing and distribution. As mentioned in Section 2, software is copyright protected only when there is human authorship, according to current legislations. If an AI coding assistant generates large code chunks, for instance entire functions, human authorship becomes questionable, unless the developer can prove, and they may need to do so in a court of law, that the AI assistant was only a tool. In a contribution written before the LLMs deluge [HQ21], responses produced by the OpenAI models of the time (GPT2 and GPT3) are explicitly mentioned as examples of AI applications whose end results would probably not qualify as copyrightable work. If the interaction between the human and the model is limited to writing prompts, the output is unlikely to show clear signs of human intellectual effort, originality, creativity, and expression, which are key requirements for copyright protection according to the authors of the study. Depending on the usage made of AI coding assistants, therefore, the code generated may not receive copyright protection. If there is no copyright protection and no other IPR applies (e.g., *sui generis* database protection), the code is rather public domain code. In this case, it is not possible to apply a proper disclaimer of liability, which is instead important to limit risks when distributing source code. In addition, the lack of licence on AI-generated code should always be considered when reusing it as part of larger projects.

The other relevant legal problem associated with the use of AI coding assistants is the risk they pose to infringe third-party IPR. AI coding assistants are trained on public code. This is certainly true for GitHub Copilot, whose training set includes the open-source repositories available on the developers' platform GitHub. As described in Section 3, the licences of open-source repositories can be very different. Some of these licences are permissive, like the MIT licence, some are weakly protective, like the LGPL licences, and some are strongly protective like the GPL licences. It is up to the user of the AI coding assistant to check that the AI-generated code does not infringe third-party IPR. Unfortunately, the developer does not really have the required information to do that, at least now, because GitHub Copilot does not provided the source(s) of the code suggestion. Microsoft, the company that owns GitHub and commercialises GitHub Copilot, has made a commitment to protect users of its AI services, including GitHub Copilot, in case of lawsuits.

Every user should read carefully the commitment to understand the extent of the legal protection provided. It should also be noted that Microsoft requires GitHub Copilot users to implement all guardrails and mitigations, such as blocking public code, to qualify for assistance in case of lawsuits. According to public statements made by GitHub, the filter that blocks public code will prevent any suggestion that contains 150 characters or more matching code in GitHub's public repositories [Salnd]. However, there are concerns about how effective these guardrails really are. In 2022, an open-source developer, Tim Davis, publicly called out GitHub Copilot because Copilot returned large chunks of his code without attribution and without LGPL license, despite having blocked the reproduction of public code [DR22]. In addition, there is no mention of how code snippets available on other websites (e.g., Stack Overflow, whose Q&A have a copyleft licence Creative Commons Share Alike 4.0 (CC BY-SA 4.0)) are filtered to avoid infringing third-party IPR. If general LLMs, like ChatGPT, are used as coding assistants there are no mitigations at all that the user can impose to filter public code. As of today, the risk to infringe third-party IPR when using an AI coding assistant to generate a substantial piece of code cannot be ruled out and, in case of IPR infraction, there is a real risk of incurring in a lawsuit. Since 2022, for instance, there is a lawsuit in California promoted by open-source developers against GitHub Copilot, Microsoft, and OpenAI, for Copilot violations of open-source licences.[10] The lawsuit is still in progress, and no decision has been made so far, but depending on the outcome, it may be the first of a series of legal proceedings against AI coding assistants.

The legal implications of AI coding assistants have remained on the background in research circles. About half of the workshop participants answered that they are not aware of the legal issues associated with the use of AI coding assistants and, quite interestingly, research institutions have been purchasing subscriptions to these services, but without establishing before institutional policies for the IPR-compliant use of these tools. The legal implications of AI coding assistants should be discussed more openly within the research community and a risk-benefit analysis should be carried out before deciding whether and how to use these tools in developing research software. Apart from the liability of developers and institutions in case of IPR infringement, there is also the long-term problem, especially relevant for large software projects, of developing a code base with unclear or problematic IPR that cannot be released under the licence originally planned for the project.

## 4.2 Artificial Intelligence Act

In 2024 the European Parliament passed the so-called *Artificial Intelligence Act* (AI Act), a comprehensive legislation for regulating AI products and services marketed in the European Union [aia24]. As of February 2025, two chapters of the AI Act have come into force: the "General Provisions" chapter that sets out the scope of the legislation and the requirement of sufficient IT literacy for the staff of the companies that deal with marketable AI systems, and the "prohibited AI Practices" chapter. The full application of the AI Act should be completed by 2027 according to the timeline made available by the European Commission.[11]

The AI Act takes a risk-based approach to AI systems. They are divided in systems with

---

[10] See the litigation webpage. An open-source developer, who is also a lawyer, involved in the lawsuit wrote an interesting blogpost on GitHub Copilot.
[11] See the website created by the European Commission for the AI Act.

Minimal Risk, Limited Risk, High Risk, and Unacceptable Risk. The systems with *Unacceptable Risk* include AI-systems undermining key human rights and promoting social scoring and biometric categorisation and control and are explicitly prohibited. The *High-risk* cases concern the use of AI systems in critical infrastructures or in a role that can be unfair to individuals or social groups. They are subject to strict obligations. *Limited Risk* AI systems have specific transparency obligations, for instance in relation to the data used for training the models. There are no specific provisions for *Minimal Risk* systems, which are currently the vast majority of the AI systems available in the EU. General-purpose AI models, like the AI coding assistants mentioned above, are also regulated in the AI Act for transparency and copyright-related issues.

The AI Acts regulates products and services that are already on the European single market or are going to be marketed there and not the AI models developed for research. Article 2.6 of the AI Act specifically excludes all AI models developed and put into service for the sole purpose of scientific research and development from the provisions of the Act. If there is no systemic risk, even general-purpose AI models that are released under a free and open-source licence and whose parameters (weights, model architecture, etc.) are made publicly available are not subject to unduly obligations. There are, however, exceptions to the rule. For instance, testing in real world conditions removes the special status of research models and the tested models need to comply with the obligations stated in the AI Act, besides ensuring to respect ethical and professional standards accepted in science and to comply with local legislations.

Given the novelty of the AI Act, many of the participants in the workshop did not pronounce themselves on the possible consequences of the new legislation for research software, but are aware that they will need to invest time to fully understand to what extent research work is exempted from the requirements set out for marketed AI products and services. Support from research organisations in navigating this complex set of AI regulations would certainly be most welcome given that legal pitfalls are around the corner. Developing AI models requires not only to write code but also to source training data and in the collection and usage of training data there are concrete risks to be sued for IPR infringement. The lawsuit brought by open-source developers against GitHub Copilot, Microsoft, and OpenAI mentioned above is only one example. Similar cases are disputed also in German courts of law. For instance, the District Court of Hamburg was called to decide on the lawful use of a copyrighted photograph in a training dataset created by a non-profit organisation (LAION v Robert Kneschke, 2024) [Qui25]. The court dismissed the lawsuit accepting the argument that the non-profit organisation, which does not host the images and does not sell the training data but makes them publicly available free of charge, had relied on the Text and Data Mining right for scientific research. The AI Act does not change this right but explicitly requires that AI models fully respect copyright dispositions. What might be acceptable in creating a training dataset might not fulfil legal requirements for IPR when the dataset is actually used to train an AI model for "real world" testing or commercial use.

## 5   Giving the Right Value to Research Software

Software is becoming an increasingly important part of research work and research output. This is true across disciplines and institutions and publicly acknowledged by funders of scientific research. An important aspect in ensuring that research software achieves its full potential is to share the source code with a proper licence. In turn, this requires to understand and comply with the legal requirements associated with developing and redistributing software. Therefore, the legal aspects of software development discussed in this paper are not a lawyer's problem, but a research software engineer's problem. Everyone developing research software needs a working knowledge of applicable regulations and potential legal issues to fully understand their duties and liability when distributing source code. Research institutions, but also associations and communities developing and supporting research software and its practitioners, should have an active role in promoting the required legal understanding to carry out the job and to make informed decisions. This training should be available to everyone developing research software, regardless of their specific job title and seniority, and should keep pace with the changing regulations.

The FLOSS community has built its identity using as a leverage the legal aspects surrounding software IPR. Although with different motivations, both the Free/Libre and the Open-Source Software communities have used the licences they developed to enable unrestricted redistribution of source code and its modifications. They have been successful in their endeavour. FLOSS software is routinely used as a building block in both commercial and non-commercial applications and its success owes much to the worldwide community of users and developers that are the driving force behind FLOSS software projects. If research software engineers want to be members of this community, as suggested by the large majority of the respondents in Karlsruhe, they need to have a good grasp of software IPR and the possibilities it opens for copyright holders. They also need to fully understand the legal value of the FLOSS licences, not just use them out of habit. Decisions about software licensing are not an administrative burden, but an opportunity to position research software within an established and thriving community. Within this community, research software engineers can find a wider audience for their work and learn here management and business strategies for sustaining long-term software projects.

A last remark worth making is that AI models, products, and services are adding a further layer of complexity to the legal aspects of research software development. Research institutions need to make decisions about the use of AI coding assistants and develop policies that minimise risks to lose copyright on the developed code and infringe third-party IPR. At the same time, if research organisations want to develop AI models and services they need to understand in detail the relevant legal framework that applies, but they also need to answer both practical and ethical questions that have a legal relevance. One of the workshop attendees in Karlsruhe, for instance, asked about the joint licensing of training data and code for AI models. Typically, research data are licensed using a Creative Commons Licence that cannot be applied to code. Vice versa research software is typically licensed using an open-source licence that cannot be applied to data. In AI models, training data, model code, and model parameters (e.g., weights) are intertwined and the question of how to make the licensing of all these components coherent and consistent is far from trivial. A quick look at the licence section on Hugging Face, a popular platform for AI models and training datasets, gives an idea of the complexity of the problem and provides examples of inconsistent choices made when licensing model and data separately. [12] How are research organisations going to address systematically this issue given the lack of established solutions? And also, are research organisations equipped to face the legal risks associated with the development of AI models that may harm human beings or discriminate against certain social groups when tested in the real world? Apart from a clear moral responsibility, there is also liability and it should be always carefully considered.

---

[12] See, for instance, this webinar given by a Microsoft corporate counsel for the Open Source Initiative.

Research software is now considered a standalone and valuable research result, but, unlike traditional research outputs like papers and textbooks, brings to the table a wider set of legal responsibilities that research institutions need to address if they want to value research software as it deserves.

# 6  Appendix: Workshop Responses

**Who is the copyright holder of research software?** (Multiple choice question)
- The research software engineer (5)
- The research institution (5)
- The research funders (0)
- It depends (12)

**What is the software licence you use most often?** (Multiple choice question)
- LGPL (1)
- MIT (7)
- Apache (1)
- GPL (8)
- Public Domain (1)
- EUPL (EU Public Licence) (1)
- Proprietary licence (2)
- Other OS licence (1)

**Does your research institution influence your choice of licence?** (Multiple choice question)
- Yes (8)
- No (16)

**Do constraints posed by your funders influence your choice of licence?** (Multiple choice question)
- Yes (12)
- No (7)

**Does your institution provide any training/advice about software licences?** (Multiple choice question)
- Yes (7)
- No (14)

**Do you feel part of the FLOSS (Free/Libre and Open Source Software) community while developing research software?** (Multiple choice question)
- Yes (15)
- No (4)

**Are you worried about infringing third-party copyright when you develop research software?**
(Scales question, range 0 (strongly disagree) to 5 (strongly agree), average rating listed)
- No, I am very confident in my understanding of software licences and I always check the licence before using third-party software in my code (avg. 3.1)
- Sometimes, but I can rely on legal advice when I need to clarify licensing issues (avg. 2.3)
- I often feel a bit lost, but I do not know to whom I should ask (avg. 3.1)
- I am honest, the licence of third-party software is not something that I check, usually (avg. 2.8)

**In case of third-party copyright infringement, who is liable?** (Multiple choice question)
- The research software engineer (2)
- The research institution (5)
- The research funders (0)
- It depends (18)

**Do you use AI coding assistants?** (Multiple choice question)
- Yes, regularly (5)
- Yes, sometimes (10)
- No, but I want to try (1)
- No, I do not trust them (7)

**If you use AI coding assistants, which one(s) do you use?** (Free answer)
- ChatGPT (multiple respondents)
- GitHub Copilot (multiple respondents)
- Claude.ai (1)
- Blablador (1)
- Google AI Studio (1)
- DeepSeek (1)
- AI assistants provided by GWDG (1)

**Does your research institution have a policy for AI coding assistants?** (Multiple choice question)
- Yes (0)
- No (12)
- Under deliberation (1)
- I do not know (8)

**Does your research institution pay a licence for using AI coding assistants?** (Multiple choice question)
- Yes (8)
- No (7)
- I do not know (4)

**Are you aware of the legal issues associated with the use of AI coding assistants?** (Multiple

choice question)
- Yes (10)
- Yes, but I do not consider them a real risk (3)
- No (9)

**Do you expect repercussions in your work due to the EU AI Act?** (Multiple choice question)
- Yes (3)
- No (8)
- Not sure (10)

**If yes, what kind?** (Free answer)
- It takes time to read [the AI Act], understand, interpret, comply.

**Acknowledgements:** I sincerely thank all the participants in the "Workshop on the EU Legal Framework for Research Software" (deRSE2025, Karlsruhe, February 2025) for their answers and insightful discussions. I am grateful to the reviewers of the paper for their comments and suggestions.